\documentclass[11pt]{article}
\usepackage{amsmath,amssymb,amsthm,verbatim}
\usepackage{color,graphics,srcltx}
\numberwithin{equation}{section}
\def\qed{\hfill\vbox{\hrule width 6 pt
\hbox{\vrule height 6 pt width 6 pt}}\medskip}
\def\IR{{\mathbb R}}   
\def\R{{\IR}}
\def\IC{{\mathbb C}} 
 
\def\cS{{\cal S}}
\def\cE{{\cal E}} 

\def\cB{{\cal B}}

\def\diag{{\rm diag}\,} 
\def\tr{{\rm tr}\,}

\def\conv{{\rm conv}}

\def\bu{{\bf u}}
\def\bv{{\bf v}}
\def\bx{{\bf x}}
\def\by{{\bf y}}

\begin{document}
\openup 1 \jot

\centerline{\bf
\large \sc On unital qubit channels}
\bigskip
\centerline{Man-Duen Choi}
\vspace*{0.015truein}
\centerline{\it Department of Mathematics, University
of Toronto,}
\baselineskip=10pt
\centerline{\it Toronto, Ontario, Canada M5S 2E4. \rm choi@math.toronto.edu}

\vspace*{10pt}
\centerline{ 
Chi-Kwong Li}
\vspace*{0.015truein}
\centerline{\it Department of Mathematics, College of William \& Mary,}
\baselineskip=10pt
\centerline{\it Williamsburg, Virginia 23187-8795, USA. \rm ckli@math.wm.edu}
\vspace*{0.21truein}
\centerline{\bf In memory of Professor Chandler Davis.}


\vspace*{0.21truein}

\begin{abstract}
A canonical form for unital qubit channels under local unitary transforms
is obtained. 
In particular, it is shown that the eigenvalues of the Choi matrix of a
unital quantum channel form a complete set of invariants of the canonical form.
It follows immediately that every unital qubit channel is the average of four
unitary channels. More generally,  a unital qubit 
channel can be expressed as the convex combination of 
unitary channels with convex coefficients $p_1, \dots, p_m$ as long as
$2(p_1, \dots, p_m)$ is majorized by the vector of eigenvalues of 
the Choi matrix of the channel. 
A unital qubit channel in the canonical form will transform
the Bloch sphere onto an ellipsoid.
We  look into the detailed structure of the natural linear maps sending the Bloch sphere 
onto a corresponding ellipsoid.
\end{abstract}
 


\section{Introduction}

The study of unital qubit channels is motivated by their importance in understanding the 
behavior of quantum systems subject to noise and errors. Unital qubit channels preserve the 
set of quantum states that form a basis for a qubit system, making them a natural class of 
channels to consider in the context of quantum error correction and fault tolerance. 
Furthermore, unital qubit channels form an important class of quantum channels in their own 
right, and studying their properties lead to better understanding of the fundamental 
properties of general quantum channels. Finally, unital qubit channels play a key role in
quantum communication, where they can limit the capacity and reliability of quantum 
communication systems. 
In fact,  there are interesting open research problems in this area. 
For example, determining the capacity of unital qubit channels is an important and 
challenging problem in quantum information theory. 
Another interesting problem is understanding the role of unital qubit 
channels in quantum state discrimination, which is a key task in many quantum information 
processing applications. One may see \cite{BK,GIN,MP,NC,VV,W} and their references
for some general background.

In this paper, we obtain some basic results on unital qubit channels. To facilitate
our discussion, let us introduce some technical background.
In the mathematical framework of quantum mechanics, quantum states are density matrices,
i.e., positive semidefinite matrices with trace one, and quantum channels (operations) 
are trace preserving completely positive linear maps $\Phi: M_n \rightarrow M_m$,
where $M_k$ denotes the set of $k\times k$ complex matrices, admitting the 
operator sum representation
\begin{equation} \label{qc-form}
\Phi(A) = \sum_{j=1}^r F_j A F_j^* \qquad \hbox{ for all } A \in M_n
\end{equation}
for some $m\times n$ matrices $F_1, \dots, F_r$ satisfying $\sum_{j=1}^r F_j^*F_j = I_n$.

For a linear map 
$\Phi: M_n \rightarrow M_m$, its Choi matrix is defined as 
$$C(\Phi) = (\Phi(E_{ij}))_{1 \le i, j \le n} 
= \sum_{i,j} E_{ij} \otimes \Phi(E_{ij}) \in M_n(M_m)\equiv M_n \otimes M_m,$$
where $\{E_{ij}: 1 \le i, j \le n\}$ is the standard basis for 
$M_n$. By a result in \cite{Choi}, a linear map $\Phi: M_n\rightarrow M_m$
is a quantum channel if and only if
$C(\Phi)$ is positive semidefinite and 
$\tr \Phi(E_{ij}) = \delta_{ij}$, the Kronecker delta.

We say that two quantum channels $\Phi, \Psi: M_n \rightarrow M_m$ are 
{\it unitarily equivalent} if there are unitary matrices $U \in M_n$ and $V \in M_m$
such that $\Phi(A) = V\Psi(UAU^*)V^*$ for all $A \in M_n$. It turns out that
this happens if and only if 
$C(\Phi) = (U^t \otimes V)C(\Psi)(U^t\otimes V)^*$, where 
$X^t$ denotes the transpose of the matrix $X$. This follows readily from
the following lemma.

\smallskip\noindent
{\bf Lemma 1.1}
{\it 
Let $\Phi: M_n\rightarrow M_k$ be a linear map.
Then for any $R, S \in M_n$,
$$\sum_{i,j} E_{ij} \otimes \Phi(RE_{ij}S) = \sum_{i,j} 
R^tE_{ij}S^t \otimes \Phi(E_{ij}).$$}

\smallskip
\it Proof. \rm 
Let $R = (r_{ij}) = \sum r_{ij} E_{ij}$ and 
$S = (s_{ij}) = \sum s_{ij} E_{ij}$. Then 
\begin{eqnarray*}
&&\sum_{i,j} E_{i,j} \otimes \Phi(RE_{ij}S) 
= \sum_{i,j,p,q}  E_{ij} \otimes \Phi(r_{pi}s_{jq} E_{pq})\\
&=& \sum_{i,j,p,q} r_{pi}s_{jq} E_{ij} \otimes\Phi(E_{p,q}) 
= \sum _{p,q} R^t E_{pq} S^t \otimes \Phi(E_{p,q}),
\end{eqnarray*}
which is the same as 
$\sum_{i,j} R^t E_{ij} S^t \otimes \Phi(E_{ij})$.
\qed

By Lemma 1.1, if $\Phi(A) = V\Psi(UAU^*)V^*$
for all $A \in M_n$, then
\begin{eqnarray*}
C(\Phi) &=& (I\otimes V) (\sum_{ij} E_{ij} \otimes \Psi(UE_{ij}U^*))(I\otimes V)^*\\
&=& (I_n\otimes V)(U^t\otimes I_m)C(\Psi)(U^t \otimes I_m)^*(I_n \otimes V)^*\\
&=& (U^t\otimes V)C(\Psi)(U^t\otimes V)^*.
\end{eqnarray*}

In Section 2, we make use of the above observation to show that if 
$\Phi: M_2 \rightarrow M_2$ is a unital qubit channel such that 
$C(\Phi)$ has eigenvalues $\lambda_1 \ge \cdots \ge \lambda_4 \ge 0$
summing up to two, then  
there are unitary matrices $U, V \in M_2$ such that 
the map $A \mapsto V^*\Phi(U^*AU)V$
has the form
\begin{equation}\label{can-form}
A \mapsto \frac{1}{2}(\lambda_1 A + \lambda_2 XAX
+ \lambda_3 YAY +  \lambda_4 ZAZ),
\end{equation}
where 
\begin{equation}\label{pauli}
X = \begin{pmatrix}0& 1 \cr 1 & 0 \cr\end{pmatrix}, \quad
Y = \begin{pmatrix}0& -i \cr i & 0 \cr\end{pmatrix},  \quad \hbox{ and } \quad
Z = \begin{pmatrix}1& 0 \cr 0 & -1 \cr\end{pmatrix}
\end{equation}
are the Pauli matrices.
Consequently, two unital qubit channels are unitarily equivalent
if and only if their Choi matrices have the same eigenvalues.

Using the results in Section 2, we show that every unital qubit channel 
$\Phi$ can be written as  the average of four unitary 
channels, i.e., maps of the form $A \mapsto UAU^*$ for some unitary 
$U$ in Section 3.  More generally, suppose 
$k$ positive eigenvalues $\lambda_1, \dots, \lambda_k$, and 
$p_1,\dots, p_m$ are positive numbers summing up to 1. 
Then there are unitary $V_1, \dots, V_m$ such that
$\Phi$ has the form $A \mapsto \sum_{j=1}^m p_j V_jAV_j^*$
$(p_1, \dots, p_m)$ is majorized by 
$\frac{1}{2}(\lambda_1, \dots, \lambda_4)$;
see Section 3 for the definition of majorization relation between two nonnegative vectors
of different sizes. 

One can identify the set of $2\times 2$ density matrices: 
$$\left\{ \frac{1}{2} \begin{pmatrix} 1 + z & x-iy \cr
x+iy & 1-z\cr\end{pmatrix}: x, y, z \in \R, \ x^2 + y^2 + z^2 \le 1\right\}$$
with the Bloch (unit) ball
$$\cB = \{(x,y,z): x, y, z \in \R, \ x^2+y^2+z^2 \le 1\}\subseteq \R^{3}.$$
\it In our discussion,  
both $\R^{1\times 3}$ and $\R^{3\times 1}$ will be referred to as $\R^3$.
\rm 
A unital qubit channel in the form (\ref{can-form}) 
will transform the Bloch ball (linearly) to an ellipsoid (including interior)
of the form
$$
\cE = \{ (d_1x, d_2y, d_3z): x,y,z\in \IR, \  
x^2 + y^2 + z^2 \le 1 \}$$
with $2(d_1, d_2, d_3) = (\lambda_1 + \lambda_2-\lambda_3 - \lambda_4,
\lambda_1 - \lambda_2+\lambda_3 - \lambda_4,
\lambda_1 - \lambda_2-\lambda_3 + \lambda_4)$; see Theorem 4.1.
Note that $\cE$ will degenerate to a point, a line segment, or  an elliptical disk if 
3, 2, or 1 of $d_1, \dots, d_3$ are zero. Using the convention $0/0 = 0$, we have 
$$\cE = \{ (x,y,z)\in \R^3: (x/d_1)^2 + (y/d_2)^2 + (z/d_3)^2 \le 1\}.$$
Section 4 is devoted to the
complete classification of quantum channels sending the Bloch ball $\cB$ onto 
an ellipsoid $\cE$ for the unital qubit channel in the canonical form 
(\ref{can-form}) in terms of $(d_1, d_2, d_3)$.

\section{A canonical form of unital qubit channels}

We begin with the following simple lemma.

\smallskip\noindent
{\bf Lemma 2.1}\it 
Let $\Phi: M_2 \rightarrow M_2$ be a unital linear map
that preserves trace and Hermitian matrices.
Then there are unitary $U, V \in M_2$ such that
the map $A \mapsto V^*\Phi(UAU^*)V$ has Choi matrix of the form 
\begin{equation}\label{lemma-abc}
\begin{pmatrix}
a & 0 & 0 & b\cr
0 & 1-a & c & 0 \cr
0 & c & 1-a & 0 \cr
b & 0 & 0 & a \cr 
\end{pmatrix} \ \hbox{ with } 
\quad 
b, c \ge 0 \ \hbox{ and } \ 2a \ge 1+b+c.\end{equation}

\smallskip
\it Proof. \rm 
Let  $P_0\in M_2$ be a rank one orthogonal projection such that 
$$\|\Phi(P_0)\|= \max\{\|\Phi(P)\|: P \hbox{ is a rank one orthogonal projection} \}.$$ 
Then there are unitary $U, V \in M_2$ such that
$P_0 = UE_{11}U^*$ and
$\Phi(P_0) = V\diag(a, 1-a)V^*$ with $\|\Phi(P_0)\| = |a|$.
So, $|a| \ge |1-a|$ implies that $a \ge 1/2$. 
Note that the map $\hat \Phi: A \mapsto V^*\Phi(UAU^*)V$ also preserves traces 
and Hermitian matrices; its Choi matrix $C(\hat \Phi)$ has the form
$$\begin{pmatrix}
a & 0 & d & b\mu_1\cr
0 & 1-a & c\mu_2 & -d \cr
\bar d & c\bar \mu_2 & 1-a & 0 \cr
b\bar \mu_1 & -\bar d & 0 & a \cr 
\end{pmatrix} \quad \hbox{ with } d, \mu_1, \mu_2 \in \IC, \ |\mu_1| = |\mu_2| = 1, \hbox{ and }
b, c \ge 0. $$
Moreover, 
$a =
\max\{\|\hat \Phi(P)\|: P \hbox{ is a rank one orthogonal projection} \}.$
We claim that $d = 0$. If not, we may let 
$w = (\cos\theta, \sin\theta|d|/d)^t$. Then the $(1,1)$ entry of
$
\hat \Phi(ww^*)$ equals 
\begin{eqnarray*}
w^* \begin{pmatrix} a & d \cr \bar d & 1-a\cr \end{pmatrix} w 
& = &a\cos^2\theta  + 2\sin\theta\cos\theta |d| + (1-a) \sin^2\theta\\
& = & a + 2\sin\theta[|d|\cos\theta - (a-1/2) \sin \theta ] > a
\end{eqnarray*}
if $\sin \theta > 0$ and  $\cos\theta > (a-1/2)\sin\theta/|d|$, 
which is a contraction.  Thus, $d = 0$ as asserted.

Let $D_1 = \diag(e^{i\theta}, 1)$ and 
$D_2 = \diag(e^{i\phi}, 1)$ such that $\mu_1 e^{i\phi}$ and 
$\mu_2 e^{-i\phi}$ have the
same argument $-\theta$. We may further replace
$\hat \Phi$ by the map
$\tilde \Phi: T \mapsto D_2(\hat \Phi(D_1 T D_1^*))D_2^*.$
Then the Choi matrix $C(\tilde \Phi)$ has the asserted form (\ref{lemma-abc})
with $b, c \ge 0$.

To prove that $2a \ge 1 + b + c$,
consider 
$P = \frac{1}{2}\begin{pmatrix}1 & 1 \cr 1 & 1 \cr\end{pmatrix}$. Then
$\tilde \Phi(P) = \frac{1}{2}\begin{pmatrix} 1 & b+c \cr b+c & 1 \cr\end{pmatrix}$ 
so that
$$a = \|\Phi(P_0)\| \ge \|\tilde \Phi(P)\| = \frac{1}{2}(1+b+c).$$
\vskip -.3in
\qed

\medskip
Applying the above lemma to unital qubit channels, we have the following.

\smallskip\noindent
{\bf Theorem 2.2} \it
Let $\Phi: M_2 \rightarrow M_2$ be a unital linear map
that preserve trace and Hermitian matrices.
Then the Choi matrix of $\Phi$ has real eigenvalues summing up to two.
If the eigenvalues are 
$\lambda_1 \ge \cdots \ge \lambda_4$, then 
there are unitary $U, V \in M_2$ such that 
the map $A \mapsto V\Phi(UAU^*)V^*$ has the form  
\begin{equation} \label{qubit-sum}
A \mapsto \frac{1}{2}(\lambda_1 A  + \lambda_2 ZAZ + \lambda_3 XAX
+ \lambda_4 YAY),
\end{equation}
where $X,Y,Z$ are the Pauli matrices described in {\rm (\ref{pauli})},
and its Choi matrix has the form
\begin{equation}\label{qubit-abc}
\frac{1}{2} {\small \begin{pmatrix} 
\lambda_1 + \lambda_2 & 0 & 0 & \lambda_1-\lambda_2 \cr
0 & \lambda_3 + \lambda_4 & \lambda_3 - \lambda_4 & 0 \cr
0 & \lambda_3-\lambda_4 & \lambda_3+\lambda_4 & 0 \cr
\lambda_1-\lambda_2 & 0 & 0 & \lambda_1+\lambda_2 \cr \end{pmatrix}}.
\end{equation}
In particular, $\Phi$ is a unital qubit channel if and only if
 $\lambda_4 \ge 0$. 
\rm

\smallskip
\it Proof. \rm By Lemma 2.1, there are unitary $U, V \in M_2$ such that
the Choi matrix of the map $A \mapsto V\Phi(U^*AU)V^*$ has the form 
(\ref{lemma-abc}).  Since $b, c \ge 0$ and $2a \ge 1+b+c$, we see that
$$(\lambda_1, \lambda_2, \lambda_3, \lambda_4)
= (a+b, a-b, 1-a+c, 1-a-c).$$ 
Thus, the map  $A \mapsto V\Phi(U^*AU)V^*$ 
has the form (\ref{qubit-sum}) with Choi matrix
(\ref{qubit-abc}).
 
 The last assertion follows from the fact that $\Phi$ is a unital qubit channel
 if and only if its Choi matrix is positive semidefinite. \qed

By Theorem 2.2, we see that  
the four eigenvalues of the Choi matrix of a unital qubit channel
form a complete set of invariants under unitary equivalence.

\smallskip\noindent
{\bf Corollary 2.3} \it
Let $\Phi, \Psi: M_2 \rightarrow M_2$ be  unital linear maps
that preserve trace and Hermitian matrices.
Then
$\Phi$  and $\Psi$ are unitarily equivalent 
if and only if  $C(\Phi)$ and $C(\Psi)$  
have the same eigenvalues (counting multiplicity).
In particular, two unital qubit channels are unitarily equivalent if and only
if their Choi matrices have the same eigenvalues.
\rm

\smallskip
\it Proof. \rm
The if part follows from Theorem 2.2.  
The  only if part follows from Lemma 1.1. \qed

Several  remarks are in order. First,
up to unitary equivalence the Choi matrix 
always has the form (\ref{qubit-abc}), and the operator sum representation
has the form (\ref{qubit-sum}).

Second,
the roles of $\lambda_1, \lambda_2, \lambda_3, \lambda_4$
in (\ref{qubit-sum}) are symmetric because 
a unital qubit channel with Choi matrix
$$
\frac{1}{2} {\small \begin{pmatrix} 
\lambda_{j_1} + \lambda_{j_2} & 0 & 0 & \lambda_{j_1}-\lambda_{j_2} \cr
0 & \lambda_{j_3} + \lambda_{j_4} & \lambda_{j_3} - \lambda_{j_4} & 0 \cr
0 & \lambda_{j_3}-\lambda_{j_4} & \lambda_{j_3}+\lambda_{j_4} & 0 \cr
\lambda_{j_1}-\lambda_{j_2} & 0 & 0 & \lambda_{j_1}+\lambda_{j_2} \cr \end{pmatrix}}
$$
for any permutation $(j_1, j_2, j_3, j_4)$, the Choi matrix has eigenvalues
$\lambda_1\ge \lambda_2\ge \lambda_3\ge \lambda_4$. By Theorem 2.2,
it is equivalent to the channel with Choi matrix (\ref{qubit-abc}).
In the following, we will often consider linear map on $M_2$
of the form $A \mapsto \mu_1 A + \mu_2 XAX + \mu_3 YAY + \mu_4 ZAZ$;
its Choi matrix will have eigenvalues
$2 \mu_1, \dots, 2\mu_4$.

Third,
under unitary similarity, the eigenvalues form a complete set of 
invariants of Hermitian matrices. In general,  two Hermitian matrices in
$M_n(M_m)$ with the same eigenvalues may not be
similar under a unitary matrix of the form
$U\otimes V$. It is interesting that the Choi matrices
of two unital qubit channels are unitarily similar if and only if
they are unitarily similar via a matrix of the form $U\otimes V$.

\smallskip\noindent
{\bf Example 2.4} {\it
 Define  $\Lambda: M_2 \rightarrow M_2$ 
by $\Lambda(A)  = \frac{1}{2}(\tr A) I$ for all $A \in M_2$.
Then $\Lambda$ is the only  unital qubit channel satisfying
$V\Lambda(UAU^*)V^*  =\Lambda(A)$ for all unitary $U, V \in M_2$.
Moreover, $\Lambda(A) = \frac{1}{4} (A + XAX +YAY+ ZAZ)$,
and  the  Choi matrix $C(\Lambda) = \frac{1}{2} I_4$    
with four equal eigenvalues.}

Using this example and the previous remark,  we can prove the following.

\smallskip\noindent
{\bf Corollary 2.5}
\it
Let $\Phi, \Psi: M_2 \rightarrow M_2$ be linear maps of the form
$$A\mapsto  \mu_1 A + \mu_2 XAX  + \mu_3 YAY + \mu_4 ZAZ
 \ \hbox{ and } \
A \mapsto \nu_1 A +\nu_2 XAX  + \nu_3 YAY +  \nu_4 ZAZ,$$
 respectively, where $\mu_j, \nu_j \in \IR$ for $j = 1, \dots, 4$,
 and $X, Y, Z$ are the Pauli matrices defined as in {\rm (\ref{pauli})}.            
Then $\Phi$ and $\Psi$ are unitarily equivalent if and only if  
$(\mu_1,\dots, \mu_4)$ is a permutation of $(\nu_1, \dots, \nu_4)$.
\rm

\smallskip
\it Proof. \rm  
Suppose $\Phi$ and $\Psi$ are unitarily equivalent.
By Lemma 1.1 and the remark after it,
there are unitary $U, V \in M_2$ such that
$C(\Phi) = (U^t\otimes V)C(\Psi)(U^t\otimes V)^*$.
So, $C(\Phi)$ and $C(\Psi)$ 
have the same eigenvalues, which are
 $2\mu_1, \dots, 2\mu_4$, and 
  $2\nu_1, \dots, 2\nu_4$, respectively.
Thus,
$(\mu_1,\dots, \mu_4)$ is a permutation of $(\nu_1, \dots, \nu_4)$.

Conversely, if
$(\mu_1, \dots, \mu_4)$ is a permutation of $(\nu_1, \dots, \nu_4)$, then
$\sum_{j=1}^4 \mu_j = \sum_{j=1}^4 \nu_j = a$.
Suppose $\Lambda$ is defined as in Example 2.4.
Then
$\hat \Phi = \Phi + (1-a)\Lambda$ and  $\hat \Psi= \Psi + (1-a)\Lambda$ 
are unital linear maps preserving trace and Hermitian matrices.
Let $\xi = (1-a)/2$.
Then
$$C(\hat \Phi) = C(\Phi) + \xi I \quad \hbox{ and } \quad 
C(\hat \Psi) = C(\Psi) + \xi I$$
have the same eigenvalues $2\mu_1 + \xi, \dots, 2\mu_4 + \xi$.
By Corollary 2.3, $\hat \Phi$ and $\hat \Psi$ are unitarily equivalent.
So, there are unitary $U, V \in M_2$  such that
$C(\hat \Phi) = (U^t\otimes V)C(\hat \Psi)(U^t\otimes V)^*.$
Clearly, we also have $C(\Phi) = (U^t\otimes V)C(\Psi)(U^t\otimes V)^*.$
Hence, $\Phi(A) = V\Psi(UAU^*)V^*$ for all $A$.
So, $\Phi$ and $\Psi$ are unitarily equivalent.
\qed

Instead of using the unitary equivalence via the Choi matrices,
we can show that a map on $M_2$ of the form  
$A \mapsto \mu_{1} A +
\mu_{2} XAX +  \mu_{3} YAY + \mu_{4} ZAZ$ 
is unitarily equivalent to a map on $M_2$ of the form 
$A \mapsto \mu_{j_1} A +
\mu_{j_2} XAX +  \mu_{j_3} YAY + \mu_{j_4} ZAZ$
for any permutation $(j_1, j_2, j_3, j_4)$ of $(1,2,3,4)$ as follows.
Let 
$H = (X+Z)/\sqrt 2$ and $H_1 = (X+Y)/\sqrt 2$.
Using the anti-commuting relations
$XY = -YX= iZ, YZ = -ZY = iX, ZX = -XZ = iY$, we have
\begin{eqnarray*}
A &\mapsto& \Psi_1(A) = H \Psi(HAH)H =  
\mu_1 A + \mu_3 ZAZ + \mu_2 XAX + \mu_4 YAY,\\
A &\mapsto& \Psi_2(A) = H_1 \Psi(H_1AH_1)H_1 =  
\mu_1 A + \mu_2 ZAZ + \mu_4 XAX + \mu_3 YAY,\\
A &\mapsto& \Psi_3(A) = Z\Psi(A)Z =  
\mu_2 A + \mu_1 ZAZ + \mu_4 XAX + \mu_3 YAY.
\end{eqnarray*}
The composition of $\Psi_2$ and $\Psi_3$ yields
$$A \mapsto \Psi_4(A) = ZH_1\Psi(H_1AH_1)H_1Z =  
\mu_2 A + \mu_1 ZAZ + \mu_3 XAX + \mu_4 YAY.$$ 
It is known that the group of permutations of $\{1,2,3,4\}$ is generated by the 
transpositions (2-cycles) $(1,2), (2,3), (3,4)$. Thus, we may get any permutation 
$(\mu_{i_1}, \dots, \mu_{i_4})$ of $(\mu_1, \dots, \mu_4)$ by
a composition of $\Psi_1, \Psi_2, \Psi_4$.

\smallskip\noindent
{\bf Remark 2.6}
We note that one can derive 
Lemma 2.1 via the correspondence between linear maps on $\IR^3$ and 
unital trace-preserving maps $\Phi: M_2\rightarrow M_2$ that preserve Hermitian matrices; see \cite{NC,Aron,RS,VV}. We gave a direct proof of Lemma 2.1 and 
used it to deduce Theorem 2.2.  Moreover, 
we showed that the matrix in (\ref{qubit-abc}) is a canonical form for the Choi matrix of 
a unital trace-preserving map $\Phi: M_2\rightarrow M_2$ that preserves Hermitian matrices, 
where $\lambda_1 \ge \cdots \ge \lambda_4$ are the eigenvalues of $C(\Phi)$. 
This led to Corollary 2.3 and Corollary 2.5, 
which provide simple tests for two unital qubit channels to be unitarily equivalent.

\section{A unital qubit channel as combination  of unitary channels}

 It is known that unitary channels are the extreme points of the set of 
  unital qubit channels; e.g., see \cite{LS}. We have the following.
 
 \smallskip\noindent
 {\bf Theorem 3.1} \it  
Let $\Phi$ be a unital qubit channel.
Then $\Phi$ is the average of four unitary channels.
\rm

\smallskip
\it Proof. \rm Note that the Choi matrix of the unitary channel
$A \mapsto VAV^*$ equals $vv^*$ with 
$v = (\alpha, -\beta, \bar \beta, \bar \alpha)^t$
if $V = \begin{pmatrix} \alpha & \bar \beta \cr -\beta & \bar \alpha
\end{pmatrix}$.
As a result, if $\alpha, \beta \in \IC$ satisfy 
$|\alpha|^2 + |\beta|^2 = 1$, then for 
$$
V_1 = \begin{pmatrix} \alpha & \bar \beta \cr -\beta & \bar \alpha \end{pmatrix},
\quad
V_2 = \begin{pmatrix} \alpha & -\bar \beta \cr \beta & \bar \alpha \end{pmatrix},
\quad
V_3 = \begin{pmatrix} \bar \alpha & \beta \cr -\bar\beta & \alpha \end{pmatrix},
\quad
V_4 = \begin{pmatrix} \bar\alpha & -\beta \cr \bar \beta & \alpha \end{pmatrix},
$$
we can let
$$
v_1 = (\alpha, -\beta, \bar \beta, \bar \alpha)^t, \quad
 v_2 = (\alpha, \beta, -\bar \beta, \bar \alpha)^t,\quad
 v_3 = (\bar\alpha, -\bar\beta,  \beta, \alpha)^t, \quad
  v_4 = (\bar\alpha, \bar\beta,  -\beta, \alpha)^t,$$
so that the Choi matrix of the unital channel 
$A \mapsto \frac{1}{4}\sum_{j=1}^4 V_j A V_j^*$ equals
$$\frac{1}{4}\sum_{j=1}^4 v_j v_j^* = \begin{pmatrix}
|\alpha|^2 & 0 & 0 & \frac{1}{2}(\alpha^2 + \bar \alpha^2) \cr
 0 & |\beta|^2 & \frac{-1}{2}(\beta^2 + \bar \beta^2) & 0 \cr
 0 & \frac{-1}{2}(\beta^2 + \bar \beta^2) & |\beta|^2 & 0 \cr
 \frac{1}{2}(\alpha^2 + \bar \alpha^2) &0 & 0 & |\alpha|^2\cr
\end{pmatrix}.$$
Now, suppose $\Phi$ is a unital channel.
By Theorem 2.2, up to a local unitary similarity transform,
a unital qubit channel has Choi matrix
$$\frac{1}{2}\begin{pmatrix} \lambda_1 + \lambda_2 & 0 & 0 & \lambda_1-\lambda_2\cr
0 & \lambda_3 + \lambda_4 & \lambda_3-\lambda_4 & 0\cr
0 & \lambda_3 - \lambda_4 & \lambda_3+\lambda_4 & 0 \cr
\lambda_1-\lambda_2 & 0 & 0 & \lambda_1+\lambda_2\cr
\end{pmatrix}
$$
with $\lambda_1, \dots, \lambda_4 \ge 0$ summing up to two.
Let $\alpha= \frac{1}{\sqrt{2}}(\sqrt{\lambda_1} + i \sqrt{\lambda_2}), 
\beta =  \frac{1}{\sqrt{2}}(
\sqrt{\lambda_4} + i \sqrt{\lambda_3})$. Then
$$|\alpha|^2 = \frac{1}{2}(\lambda_1 + \lambda_2), 
\quad \alpha^2+\bar \alpha^2
= \lambda_1-\lambda_2, \quad
|\beta|^2 = \frac{1}{2}(\lambda_3 + \lambda_4), \quad 
\beta^2+\bar \beta^2
= \lambda_4-\lambda_3.$$
As a result, up to a unitary similarity transform,
$\Phi$ has the form $A \mapsto \frac{1}{4} \sum_{j=1}^4 V_j A V_j^*$
as defined above. 
\qed

The set $\cS$ of unital qubit channels is linearly 
 isomorphic to the compact convex set of Choi matrices of unital qubit channels,
 that are $4\times 4$ positive semidefinite matrices in block form 
 $(C_{ij})_{1 \le i, j \le 2}$ with $C_{11}, C_{12}, C_{21}, C_{22} \in M_2$
 satisfying $C_{11}+C_{22} = I_2$, $\tr C_{11} = \tr C_{22} = 1$ 
 and $\tr C_{12} = \tr C_{21} = 0$. Thus, the set of Choi matrices
 has real affine dimension 9, and it is known that the Choi matrices
 of unitary channels are extreme points of the convex set; see \cite{LS}. 
 By the Caratheodory theorem, every element 
 in $\cS$ is a convex combination of no more than 
 10 extreme points. It is remarkable that
Theorem 3.1 ensures that every element $\Phi\in \cS$ 
can be written as the average of four extreme points.

\medskip
In the following, we will determine all the possible convex coefficients
$p_1,\dots,p_m$ so that $\Phi = \sum_{j=1}^m p_j \Psi_j$
for some unitary channels $\Psi_1,\dots, \Psi_m$.  Theorem 3.2 asserts
that the set of such $(p_1, \dots, p_m)$ can be completely determined by 
the eigenvalues of the Choi matrix of $\Phi$.
To achieve this, we need the concept of majorization.

\medskip
Recall that a vector  $u\in \IR^{m}$ is {\it  majorized} by another 
vector $v\in \IR^{m}$,
denoted by $u \prec v$, if the sum of the $k$ largest entries of $u$ is not
larger than that of $v$ for $k = 1, \dots, m$, with equality holding for 
$k = m$. We extend the notion of majorization to nonnegative vectors of different sizes
as follows. Suppose the sum of entries of the nonnegative vectors 
$u \in \IR^m$ and $v \in \IR^{n}$  are the same. 
We can change $u, v$ to $\tilde u, \tilde v$ by adding zero entries
to the vector with lower dimension. Then we say that $u \prec v$ if 
$\tilde u \prec \tilde v$.

Our main theorem is the following.

\smallskip\noindent
{\bf Theorem 3.2} \it 
Let
$\Phi: M_2\rightarrow M_2$ be a unital qubit quantum channel
such that the Choi matrix of $\Phi$  has $k$ positive eigenvalues
$\lambda_1 \ge \cdots \ge \lambda_k$.
Suppose $p_1, \dots, p_m$ are positive real numbers summing up to one.
Then there are unitary matrices $V_1, \dots, V_m \in M_2$ such that
\begin{equation} \label{p-form}
\Phi(A) = \mu_1 V_1 A V_1^* + \cdots + \mu_m V_m A V_m^*  \qquad \hbox{ for all } A \in M_2
\end{equation}
if and only if 
$(\mu_1, \dots, \mu_m) \prec 
\frac{1}{2}(\lambda_1, \dots, \lambda_k).$
\rm

\smallskip

From Theorem 3.2, one can deduce the following corollary from which Theorem 3.1 will follow.
 
 \smallskip
 \noindent
 {\bf Corollary 3.3} \it Suppose $\Phi$ is a unital qubit channel
 such that $C(\Phi)$ has $k$ positive eigenvalues.
 Then $\Phi$ can be written as the average of $m$ unital 
 channels for any positive integer $m$ satisfying $k \le m$.
\rm
 
 \smallskip
 \it Proof. \rm If $C(\Phi)$ has eigenvalues $\lambda_1, \dots, \lambda_k$, then
 $(1/m, \dots, 1/m) \prec \frac{1}{2}(\lambda_1, \dots, \lambda_k)$.
 So, $C(\Phi)$ is the average of $m$ unitary channels by Theorem 3.2.
 \qed

\medskip
To prove Theorem 3.2, we first obtain 
some auxiliary results, which are of independent interest.

\medskip\noindent
{\bf Lemma 3.4} \it 
Let $\theta, \eta_1, \eta_2, \nu_1, \nu_2$ be nonnegative numbers such that
$\eta_1 \ge \nu_1 \ge \nu_2 \ge \eta_2$
and $\eta_1 + \eta_2 = \nu_1 + \nu_2 = d$.
Then there are $\theta_1, \theta_2\in [0,2\pi)$ with
$\nu_1 e^{i\theta_1} + \nu_2 e^{i\theta_2} = \eta_1+\eta_2e^{i\theta}$.
\rm

\smallskip
\it Proof. \rm 
By the given conditions,
$$\nu_1 - \nu_2 \le \eta_1 - \eta_2
\le 
|\eta_1 + \eta_2 e^{i\theta}|
\le \eta_1 + \eta_2 = \nu_1 + \nu_2.$$
Thus, there is $\phi \in [0, 2\pi)$ such that
$|\nu_1 + e^{i\phi} \nu_2 | = |\eta_1 + \eta_2 e^{i\theta}|$.
Hence, there is $\theta_1\in \IR$ such that
$e^{i\theta_1}(\nu_1 + e^{i\phi}\nu_2) = \eta_1 + \eta_2 e^{i\theta}$. 
Let $\theta_2 \equiv \phi + \theta_1$ (mod $2\pi$). The result follows.
\qed

\smallskip
\noindent
{\bf Corollary 3.5} \it
Suppose $\Psi(A) = \eta_1 V_1AV_1^* + \eta_2 V_2AV_2^*$
for all $A \in M_2$, where $\eta_1 \ge \eta_2 \ge 0$,
$V_1, V_2 \in M_2$ are unitary matrices. 
Then for any  $\nu_1, \nu_2$ such that $\eta_1 \ge \nu_1 \ge \nu_2 \ge \eta_2$
with $\eta_1 + \eta_2 = \nu_1 + \nu_2$,  there 
are unitary matrices $U_1, U_2 \in M_2$
such that
$\Psi(A) = \nu_1U_1AU_1^*+ \nu_2 U_2AU_2^*$ for all $A \in M_2$.
\rm

\smallskip
\it Proof. \rm 
Consider the special case when $V_1 = I_2$ and $V_2 =D = \diag(1, e^{i\theta})$.
Define $\Psi(A) = \eta_1 A + \eta_2 DAD^*$. It follows that 
$$
C(\Psi) = \eta_1 
\begin{pmatrix} 
1 & 0 & 0 & 1\cr
0 & 0 & 0 & 0\cr
0 & 0 & 0 & 0\cr
1 & 0 & 0 & 1\cr
\end{pmatrix} + \eta_2
\begin{pmatrix} 
1 & 0 & 0 & e^{-i\theta}\cr
0 & 0 & 0 & 0\cr
0 & 0 & 0 & 0\cr
e^{i\theta} & 0 & 0 & 1\cr\end{pmatrix}.
$$
By Lemma 3.4, there are $\theta_1, \theta_2 \in [0, 2\pi)$ such that  
$\eta_1 + \eta_2 e^{i\theta} = \nu_1 e^{i\theta_1} + \nu_2 e^{i\theta_2}$.
Let $\tilde U_1 = \diag(1, e^{i\theta_1}), \tilde U_2 = \diag(1, e^{i\theta_2})$.
Then $\Psi(A) = \nu_1 \tilde U_1 A \tilde U_1^* + \nu_2 \tilde U_2 A\tilde U_2^*$.

Next, consider general unitary $V_1, V_2 \in M_2$. Then $V_1^*V_2 = \alpha WDW^*$
with a complex unit $\alpha$, a unitary $W \in M_2$ and 
$D = \diag(1, e^{i\theta})$. Then 
\begin{eqnarray*}
\Psi(A) &=&  \eta_1 V_1AV_1^* + \eta_2 V_2AV_2^* = 
\eta_1 V_1AV_1^* + \eta_2 V_1(V_1^*V_2 A_2 V_2^* V_1)V_1^* \\
& =& \eta_1 V_1AV_1^* + \eta_2 V_1(\alpha WDW^* A\bar \alpha WD^*W^*)V_1^*
= V_1W\Phi(W^*AW)W^*V_1^*
\end{eqnarray*} 
with
$\Phi(X) = \eta_1 X + \eta_2 DXD^*$ for $X \in M_2$.
By the special case, there are diagonal unitary matrices 
$\tilde U_1, \tilde U_2 \in M_2$
such that
$$\Phi(W^*AW) = \eta_1 (W^*AW) + \eta_2 D(W^*AW)D^*
= \nu_1 \tilde U_1 (W^*AW) \tilde U_1^* + \nu_2 \tilde U_2(W^*AW)\tilde U_2^*.$$
Let 
$U_j = V_1W \tilde U_j W^*$ for $j = 1,2$. Then 
$\Psi(A) = \nu_1 U_1 AU_1^* + \nu_2U_2AU_2^*$
for all $A \in M_2$. \qed

\smallskip\noindent
{\bf Lemma 3.6} \it
Let $m \ge 2$ and $\bu = (u_1, \dots, u_m), \bv = (v_1, \dots, v_m) \in \IR^m$ with 
entries arranged in descending order, and $\bu \prec \bv$. Then one of the 
following holds.
\begin{itemize}
\item[{\rm (1)}] There is $i \in \{1, \dots, m\}$ such that
$u_i = v_i$.
\item[{\rm (2)}] There is $j \in \{1, \dots, m-1\}$ with 
$v_j > u_j \ge u_{j+1} > v_{j+1}$. If
$\delta = \min\{v_j-u_j, u_{j+1}-v_{j+1}\}$,  
then the vector $\tilde \bv$ obtained from $\bv$
by replacing the pair of entries
$(v_j, v_{j+1})$ to $(v_j-\delta, v_{j+1} + \delta)$ will satisfy
$\bu\prec \tilde \bv \prec \bv$, and $\bu$ and $\tilde \bv$ will have the same
value at the $j$th or $(j+1)$st position.
\end{itemize}

\smallskip

\it Proof. \rm Direct verification. \qed

In general, 
suppose $\by = (y_1, \dots, y_m)$ 
with $y_1 \ge  \cdots \ge y_m$,
and  $\bx = (x_1, \dots, x_m)$ 
is obtained from $\by$ by changing a pair of entries
$(y_p, y_{q})$ to $(x_p, x_q) = (y_p-\delta, y_{q} + \delta)$ with 
$\delta \in (0, (y_p-y_{q})/2)$ for some $p < q$ and $y_q - y_p > 0$.
We say that $\bx$ is obtained from $\by$ by a pinching.
By Lemma 3.6, if 
$\bx, \by \in \IR^m$ satisfy $\bx \prec \by$, we may remove
the entries $x_\ell, y_\ell$ whenever $x_\ell = y_\ell$, and 
apply (2) to get a vectors $\tilde \by$ such that  the pair of vectors
$\bx$ and $\tilde \by$ have one more common entry than the pair of vectors
$\bx$ and $\by$. 
Thus, one can change $\by$ to  $\bx$  by at most $m-1$ pinchings.

\smallskip\noindent
{\bf Corollary 3.7} \it
Suppose $\Phi$ is a qubit channel having the form 
\begin{equation} \label{nu-form}
A \mapsto \sum_{j=1}^k \nu_j U_j A U_j^*
\end{equation}
for some nonnegative real numbers  $\nu_1, \dots, \nu_k$ summing up to 1, and 
unitary $U_1, \dots, U_k\in M_2$.
If 
$(\mu_1, \dots, \mu_m) \prec (\nu_1, \dots, \nu_k)$,
then $\Phi$ admits a representation of the form $A \mapsto \sum_{j=1}^m \mu_j V_jAV_j^*$
for some unitary $V_1, \dots, V_m \in M_2$.
\rm

\smallskip

\it Proof. \rm We may assume that $m = k$ by adding zeros to 
the vector with lower dimension. 
Assume $(\mu_1, \dots, \mu_m) \ne (\nu_1, \dots, \nu_m)$
to avoid trivial consideration. Furthermore, we may assume that the entries
of the two vectors are arranged in descending order. 
Suppose $(\mu_1, \dots, \mu_m)$ is obtained from  $(\nu_1, \dots, \nu_m)$ 
by a pinching, say, $(\mu_p, \mu_q) = (\nu_p-\delta, \nu_q + \delta)$
with $\delta \in (0, (\nu_q-\nu_p)/2)$ for some $p < q$.
By Lemma 3.5, we can replace the expression 
$\nu_p U_pA U_p^* + \nu_{q} U_{q} AU_{q}^*$ in ({\ref{nu-form})
by $\mu_p V_pAV_p^* + \mu_{q} V_{q} AV_{q}^*$ 
for some unitary matrices  $V_p, V_{q} \in M_2$.
By the remark after Lemma 3.6, we can reduce 
$(\nu_1, \dots, \nu_m)$ by at most $m-1$ pinchings.
Thus, we can repeat the above argument for at most $m-1$ times
to get the conclusion.
 \qed

Now we are ready to present the following.

\smallskip\noindent
\bf  Proof of Theorem 3.2. \rm
Suppose $\Psi(A) = \sum_{j=1}^\ell V_j A V_j^*$ for all $A \in M_n$.
Then 
$C(\Psi) = \sum_{j=1}^\ell v_j v_j^*$, where for $j = 1, \dots, \ell$,
$v_j \in \IC^{mn}$
has the first $m$ entries equal to the first column of $V_j$, the next
$m$ entries equal to the second column of $V_j$, etc.
Suppose $R= [v_1 | \cdots | v_\ell]$. Then 
$C(\Psi) = RR^*$ has the same nonzero eigenvalues as $R^*R \in M_\ell$,
which is a Hermitian matrix with diagonal entries
$$(v_1^*v_1, \dots, v_\ell^* v_\ell) = 
(\tr(V_1^*V_1), \dots, \tr(V_\ell^*V_\ell)).$$
Then $\ell \ge k$ and the majorization relation holds by the result in 
\cite{S}.

For the sufficiency,  we see that $\Phi$ can be written in the form 
$A \mapsto \frac{1}{2}\sum_{j=1}^k \lambda_j U_jAU_j^*$  by Theorem 2.2.
By Corollary 3.7,
If  $(\mu_1, \dots, \mu_m) \prec 
\frac{1}{2}(\lambda_1, \dots, \lambda_k),$ there are unitary $V_1, \dots, V_m \in 
M_n$  such that $\Phi$ can be written in  the form (\ref{p-form}).
\qed

We note that Corollary 3.5,
Corollary 3.7, and Corollary 3.3 were obtained as Lemma 1.1, Theorem 1.2, and 
Corollary 1.4
in \cite{MP}, where the proofs were done using the idea in \cite{CW} as indicated by 
the authors. 

We conclude this section with some comments on the unusual 
convexity features of the set $\cS$ of unital qubit channels,
whose extreme points are unitary channels.

By Theorem 3.2, if $\Phi \in \cS$ such that 
$C(\Phi)$ has rank $k$ with positive eigenvalues
$\lambda_1 \ge \cdots \ge \lambda_k$, 
then $\Phi$ can be written as a convex combination of unitary channels
$A \mapsto \sum_{j=1}^m \mu_j U_jAU_j^*$ as long as
$(\mu_1, \dots, \mu_m) \prec 
\frac{1}{2}(\lambda_1, \dots, \lambda_k)$.
Consequently, the four eigenvalues of $C(\Phi)$ 
can be characterized as the supremum of set 
of four numbers $p_1 \ge \cdots \ge p_4$
such that $\Phi$ admits a representation of the form 
$A \mapsto \frac{1}{2}(p_1 V_1AV_1^* + \cdots + p_4 V_4AV_4^*)$
for some unitary matrices $V_1, \dots, V_4 \in M_2$.
Of course, 
$(\lambda_1, \cdots, \lambda_4)$ 
can also be characterized as $(\tr V_1^*V_1, \dots, \tr V_4^*V_4)$
such that $\Phi$ admits a representation of the form 
$A \mapsto  V_1AV_1^* + \cdots + V_4AV_4^*$
for some matrices $V_1, \dots, V_4 \in M_2$ 
satisfying $\tr V_i^*V_j = 0$ for all $1 \le i, j \le 4$.
If $\Phi$ is a unitary channel so that 
$\Phi$ is a unital qubit channel so that $C(\Phi)$ has eigenvalues
$1,0,0,0$, then $\Phi$ can be written as $\sum_{j=1}^m \mu_j U_jAU_j^*$ for
any choice of convex coefficients $\mu_1, \dots, \mu_m$.
In some sense, the eigenvalues of $C(\Phi)$ can be viewed as a measure on 
how close $\Phi$ is to a unitary channel, an extreme point
of $\cS$.

\section{Unital qubit channels and the Bloch ball}

A  unital qubit channel
$\Phi: M_2 \rightarrow M_2$ in the canonical form 
(\ref{can-form}) is determined by the conditions
\begin{equation}\label{eq4.0}
\Phi(I_2) = I_2,  \quad
\Phi(X) = d_1X, \quad \Phi(Y) = d_2Y,
\quad \Phi(Z) = d_3Z,\end{equation} 
where $d_1 = (\lambda_1 +\lambda_2- \lambda_3 - \lambda_4)/2$,
$d_2 = (\lambda_1 - \lambda_2 + \lambda_3 -\lambda_4)/2$, and
$d_3 = (\lambda_1 - \lambda_2 - \lambda_3 +\lambda_4)/2$.
The map $\Phi$ will transform  the Bloch (unit) ball to the ellipsoid $\cE$ specified by
$(d_1,d_2,d_3)$:
$$\ \{ (d_1x, d_2y, d_3 z): x,y,z \in \IR, x^2+y^2+z^2 \le 1\}
$$
$$= \{(x,y,z)\in \IR^3: (x/d_1)^2 + (y/d_2)^2 + (z/d_3)^2\le 1\}.$$
If $\Phi$ is defined as in  (\ref{eq4.0}), then  
$$\Phi(E_{11}) = \Phi(I) +\Phi(Z) = 
\frac{1}{2} \begin{pmatrix} 
1+ d_3 &0\cr
0 & 1-d_3 \cr \end{pmatrix}, \
\Phi(E_{22}) = I-\Phi(E_{11}) = 
\frac{1}{2} \begin{pmatrix} 
1- d_3 &0\cr
0 & 1+d_3 \cr \end{pmatrix},$$
and 
$$\Phi(E_{21})^t = \Phi(E_{12}) = \frac{1}{2} (\Phi(X) + \Phi(iY)) =  
\frac{1}{2} \begin{pmatrix} 
0 & d_1 + d_2\cr d_1-d_2 & 0 \cr\end{pmatrix}.$$
By these observations and  Theorem 2.2,  we have the following. 

\smallskip\noindent
{\bf Theorem 4.1} \it
Let $\Psi: M_2 \rightarrow M_2$ be a unital trace-preserving linear map that preserves
 Hermitian matrices. Then $\Psi$ is unitarily equivalent to $\Phi$ of the form 
  {\rm (\ref{eq4.0})} sending the Bloch sphere onto the ellipsoid specified by 
  $(d_1,d_2,d_3)$. The Choi matrix $C(\Phi)$ has the form
\begin{equation}\label{Phi}
\frac{1}{2} \begin{pmatrix} 
1+ d_3 &0& 0 & d_1 + d_2\cr
0 & 1-d_3 & d_1 - d_2 & 0 \cr
0 & d_1-d_2 & 1-d_3 & 0 \cr
d_1+d_2 & 0 & 0 & 1+d_3\cr
\end{pmatrix}
\end{equation}
with eigenvalues 
$\frac{1}{2}(1+d_1+d_2+d_3), 
\frac{1}{2}(1+d_3-d_1-d_2), \frac{1}{2}(1-d_3+d_1-d_2),
\frac{1}{2}(1-d_3-d_1+d_2)$.
\rm

\medskip

One can now connect the results in Sections 2 and 3 to the map
$(x,y,z) \mapsto (d_1x, d_2y, d_3z)$.
For example, by Corollary 2.5,
we have the following.

\smallskip\noindent
{\bf Theorem 4.2}
\it Let $\Phi, \Psi: M_2\rightarrow M_2$ be linear map
such that $\Phi$ is defined by {\rm (\ref{eq4.0})} and
$\Psi$ is defined by
$$
\Psi(I_2) = I_2,  \quad
\Psi(X) = \tilde d_1X, \quad \Psi(Y) = \tilde d_2Y,
\quad \Psi(Z) = \tilde d_3Z.$$
Then the following conditions are equivalent.

\medskip
{\rm (a)}   
$\Phi$ and $\Psi$ are unitarily equivalent.

\medskip
{\rm (b)} $(\tilde d_1+\tilde d_2+\tilde d_3, 
\tilde d_1-\tilde d_2-\tilde d_2, -\tilde d_1+\tilde d_2-\tilde d_3, 
-\tilde d_1-\tilde d_2+\tilde d_3)$ is a 
permutation of 

\qquad $(d_1+d_2+d_3, d_1-d_2-d_3, -d_1+d_2-d_3, -d_1-d_2+d_3)$.

\medskip
{\rm (c)}  
$(|\tilde d_1|, |\tilde d_2|, |\tilde d_3|)$ 
is a permutation of $(|d_1|, |d_2|, |d_3|)$, and 
$\tilde d_1\tilde d_2\tilde d_3 = d_1d_2d_3$.
\rm

\smallskip

\it Proof. \rm  
By Corollary  2.3,
(a) holds if and only if $C(\Phi)$ and $C(\Psi)$ have the same
eigenvalues,  equivalently, condition (b) holds by (\ref{Phi}). 

To prove the equivalence of (b) and (c), let 
$T = \frac{1}{2} \begin{pmatrix} 
1 & 1 & -1 & -1 \cr
1 & -1 & 1 & -1 \cr
1 & -1  & -1 & 1 \cr\end{pmatrix}$.
Then $TT^* = I_3$ and $T^*T = I_4 - J_4/4$, where $J_4 \in M_4$
has all entries equal to 1.  Moreover, for any permutation matrix $P \in M_4$
and $PT^*T = T^*TP$.
Consequently, the map 
$X \mapsto TXT^*$ defines a group homomorphism  from the group of permutation
matrices in $M_4$ to the group of the 24 matrices $Q_1, \dots, Q_{24} \in M_3$, 
where  each $Q_j$ is the product of a permutation
matrix in $M_3$ and a diagonal orthogonal matrix with determinant 1.

Note that
$(d_1+d_2+d_3, d_1-d_2-d_3, -d_1+d_2-d_3, -d_1-d_2+d_3)
= 2(d_1,d_2, d_3)T.$
Thus, if (b) holds, then there is a permutation matrix $P \in M_4$
such that
$(\tilde d_1,\tilde d_2,\tilde d_3)T = (d_1,d_2,d_3)TP$.
Hence,
$(\tilde d_1, \tilde d_2, \tilde d_3) = (d_1,d_2,d_3)TPT^* = (d_1, d_2, d_3) Q_j$
for some $j = 1, \dots, 24$, i.e., condition (c) holds.

If (c) holds, then $(\tilde d_1, \tilde d_2, \tilde d_3) = (d_1, d_2, d_3)Q$
for some $Q \in \{Q_j: 1 \le j \le 24\}$ 
so that $(\tilde d_1, \tilde d_2, \tilde d_3)T =
(d_1, d_2, d_3)(TPT^*)T = (d_1, d_2, d_3)T(T^*T)P = (d_1,d_2,d_3)TP$
for some permutation $P \in M_4$. Thus, condition (b) holds.
\qed

\medskip
Note that if  $\Phi: M_2 \rightarrow M_2$ satisfying (\ref{eq4.0})  
is a unital qubit channel, then it will sends the set of density matrices
back to itself. Hence, $|d_j| \le 1$ for $i = 1, \dots, 3$.
However, the converse is not true in general.
For example, if we let $(d_1,d_2,d_3) = (1,1,0)$, then
\begin{equation}\label{map}
\Phi: \frac{1}{2}\begin{pmatrix}1+z & x-iy\cr x+iy& 1-z\cr\end{pmatrix} \mapsto 
\frac{1}{2}\begin{pmatrix}1 & x-iy\cr x+iy& 1\cr\end{pmatrix}
\quad \hbox{ with } \quad  C(\Phi) = \frac{1}{2}\begin{pmatrix}
1 & 0 & 0 & 2 \cr
0 & 1 & 0 & 0 \cr
0 & 0 & 1 & 0 \cr
2 & 0 & 0 & 1 \cr\end{pmatrix}
\end{equation}
is not a quantum channel because $C(\Phi)$
is not positive semidefinite.

We have shown that up to unitarily equivalence,
every unital qubit channel can be written as a convex 
combination of the following four maps
$$A \mapsto A, \qquad A \mapsto XAX, \qquad  A \mapsto YAY, \qquad 
A \mapsto ZAZ.$$
The corresponding linear transformations of the Bloch balls are 
$$(x,y,z) \mapsto (x,y,z), \ (x,y,z) \mapsto (x,-y,-z), \
(x,y,z) \mapsto (-x,y,-z), \ (x,y,z) \mapsto (-x-y,z),$$
which are the identity map, the rotation of $\pi$ of the 
$(y,z)$-plane, $(x,z)$-plane, and $(x,y)$-plane, respectively.
Equivalently, $(d_1, d_2, d_3)$ 
is a convex combination of vectors 
$$(1,1,1), \  (1,-1,-1), \ (-1,1,-1), \ (-1,-1,1).$$
As a result, $\Phi: M_2\rightarrow M_2$ in the canonical form
(\ref{can-form}) transforming the Bloch ball by the map
$(x,y,z) \mapsto (d_1x, d_2y, d_3z)$
is a unital qubit channel if and only if  the 
eigenvalues of $C(\Phi)$ are nonnegative, i.e.,
\begin{equation}\label{d123}
1+d_1+d_2+d_3, 
1+d_3-d_1-d_2,  1-d_3+d_1-d_2,  
1-d_3-d_1+d_2 \in [0, \infty).
\end{equation}
One can 
determine the set of extreme points of the compact convex set 
$(d_1, d_2, d_3) \in \IR^3$ satisfying (\ref{d123}).
Every extreme point of the set must attain equality for 
at least three of the linear inequality constraints.
A direct computation shows that there are four extreme points,
namely, 
$$(1,1,1), (1,-1,-1), (-1,1,-1), (-1,-1,1),$$ 
and the set of real vectors $(d_1,d_2,d_3)$ satisfying (\ref{d123})
is the convex hull of these four points, which is a
regular tetrahedron.
These observations can be used to determine a linear map $\Phi: M_2\rightarrow M_2$
satisfying (\ref{eq4.0}) is a unital qubit channel as shown in the following.

\smallskip
\noindent
{\bf Theorem 4.3} \it
Suppose $\Phi: M_2 \rightarrow M_2$ is a linear map as defined by 
{\rm (\ref{eq4.0})}. Then $\Phi$ is a unital qubit channel if and only if 
any one of the following holds.
\begin{itemize} 
\item[
{\rm (a)}] $(d_1, d_2, d_3)$ belongs to the 
regular tetrahedron with vertices 
$$(1,1,1), (1,-1,-1), (-1,1,-1), (-1,-1,1).$$

\item[
{\rm (b)}] The vector
$(1+d_1+d_2+d_3, 
1+d_1-d_2-d_3, 1-d_1+d_2-d_3, 1-d_1-d_2+d_3)$ is  nonnegative.
\end{itemize}
\rm

\medskip
In Theorem 4.3, (a) and (b) establish a correspondence between 
the regular tetrahedron in $\IR^{3}$ with vertices 
$(1,1,1), (1,-1,-1), (-1,1,-1), (-1,-1,1)$, and the  convex set 
$$\left\{(\lambda_1,\dots, \lambda_4): \lambda_1, \dots, \lambda_4 \ge 0,
\sum_{j=1}^4 \lambda_j = 2\right\}.$$ 
Mathematically, it is simple to verify the correspondence.
Nevertheless, it is physically astonishing that the
unital qubit channels serve as the natural link between these two convex sets.
Namely, each quantum channel $\Phi$  determines four eigenvalues 
$\lambda_1, \dots, \lambda_4$ of $C(\Phi)$ as well as the shrinking
effects $(d_1, d_2, d_3)$ of the Block sphere in $\IR^3$.
For example, if $(d_1, d_2, d_3) = (1,1,0)$, i.e., the Bloch sphere
is transformed to the unit disk on the $(x,y)$-plane, then it does not
correspond to a unital qubit channel. Also, if $(d_1, d_2, d_3) = (1,1,-1)$,
i.e., the Bloch sphere is transformed to itself by a reflection about the
$(x,y)$-plane, then it does not correspond to a unital qubit channel.

In fact, it not hard to show that for a nonngative number
$\gamma$ the map $(x,y,z) \mapsto \gamma(x,y,-z)$
corresponds to a unital qubit channel if and only if $\gamma \in [0, 1/3]$.
It turns out that this and other limitations on admissible scalings 
of the Bloch sphere corresponding to unital qubit maps can be deduced from 
the next theorem, where the assumption  $d_1 \ge d_2 \ge |d_3|$
is imposed without loss of generality by virtue of Theorem 4.2 (c).

\smallskip\noindent
{\bf Theorem 4.4} \it
Suppose $\Phi: M_2 \rightarrow M_2$ is a linear map as defined by {\rm (\ref{eq4.0})}
with  $d_1 \ge d_2 \ge |d_3|$. Then $\Phi$ is 
a unital qubit channel if and only if $1+d_3 \ge d_1 + d_2$, equivalently,
$(d_1, d_2, d_3) \in 
\conv\{(0,0,0), (1,1, 1), (1,0,0), (1,1,-1)/3\}.$

\smallskip

\it Proof. \rm
By Theorem 4.3, under the assumption that $d_1 \ge d_2 \ge |d_3|$,
condition (b.2) reduces to $1+d_3 \ge d_1 + d_2$.
The convex set 
$$S = \{(d_1, d_2, d_3) \in \IR^3: d_1 \ge d_2 \ge |d_3|, 1+d_3 \ge d_1 + d_2\}$$
is defined by the four inequalities $d_1 \ge d_2, d_2 \ge d_3, d_2 \ge -d_3$,
and $1+d_3 \ge d_1+d_2$; an extreme point of the set $S$ must attain 
at least three of these inequalities. Thus the  extreme
points of  $S$ are $(0,0,0), (1,1,1), (1,0,0), (1,1,-1)/3$, and
$$S = \conv\{(0,0,0), (1,1,1), (1,0,0), (1,1,-1)/3\}.$$
\vskip -.3in
\qed

\medskip
By Theorem 4.4, $\Phi$ is a unital qubit channel
if and only if 
the Bloch ball is transformed by a convex combination of the maps
$\psi_1: (x,y,z) \mapsto (0,0,0)$, 
$\psi_2: (x,y,z) \mapsto (x,y,z)$,
$\psi_3: (x,y,z) \mapsto (x,0,0)$,  
$\psi_4: (x,y,z) \mapsto (0,0,-z).$
These maps on the Bloch sphere (the boundary of the Bloch ball) 
have the following effects,
respectively:
\begin{itemize}
\item shrinking the sphere to  the origin, 
\item leaving the sphere invariant, 
\item shrinking the Bloch sphere to
the line segment joining $(-1,0,0)$ to $(1,0,0)$,
\item shrinking the Bloch sphere by a factor of $1/3$
followed by a refection over the $(x,y)$-plane.
\end{itemize}
In terms of  $\Phi$, it means that every unital qubit channel 
can be written as a convex combination of
$\Psi_1, \dots, \Psi_4$ defined, respectively,  by
$$
 A \mapsto \frac{1}{2}(\tr A)I_2, \quad 
A \mapsto A, \quad  A \mapsto \frac{1}{2}(A + XAX), 
\quad \hbox{ and } \quad 
A \mapsto \frac{1}{3}(A+XAX+YAY).
$$

Using Theorem 4.2 (c),
one may impose other assumptions on the scaling factors $d_1, d_2, d_3$
of the Bloch sphere, and deduce additional results on  unital qubit channels. 
For example,  we may assume that $d_1 \ge d_2 \ge d_3$ and $d_2 \ge 0$.
In such a case, $(x,y,z) \mapsto (d_1x, d_2y, d_3z)$ will correspond to
a unital qubit channel if and only if $d_1 \ge d_2 \ge d_3$, $d_2 \ge 0$,
and $1+d_3 \ge d_1 + d_2$, i.e., 
$(d_1, d_2, d_3) \in \conv\{(0,0,0), (1,1,1),(1,0,0), (0,0,-1)\}$.
Thus, a unital qubit channel can be written as a convex
combination of the maps 
$$ A \mapsto \frac{1}{2}(\tr A)I_2, \quad 
A \mapsto A, \quad  A \mapsto \frac{1}{2}(A + XAX), 
\quad \hbox{ and } \
A \mapsto \frac{1}{2}(XAX+YAY).$$

\section{Additional remarks and further research}

One referee pointed out that the Choi matrix of a unital quantum channel corresponds to 
a maximally entangled 2-qubit state after normalization, as noted in our paper. 
Specifically, if $\rho = (\rho_{ij}) \in M_2(M_2)$ is a density matrix with $\rho_{11} + 
\rho_{22} = I_2/2$, then $\rho$ represents a maximally entangled 2-qubit state. Corollary 
2.3 further states that two maximally entangled 2-qubit states $\rho_1$ and $\rho_2$ 
in $M_4$ are unitarily similar if and only if there exist unitary matrices $U, V \in M_2$ 
such that $(U\otimes V)\rho_1 (U\otimes V)^* = \rho_2$.
It was also suggested that one may study unital qutrit or qudit channels. However,
qutrit and qudit channels may not be mixed unitary so that our techniques do not apply.
New techniques will be required to study these  problems.

Another referee pointed out that 
the matrix in (\ref{lemma-abc}) can be interpreted as 
an $X$-state in $M_4$ after normalization, if it is positive semidefinite. 
In \cite{Caban}, the authors identified 2-qubit states that can be 
reduced to such states under a certain equivalence relation; 
see Theorem 1 and Section 2.3 in \cite{Caban}.
Our study may have other connections to the study of $X$-states, 
e.g., see  \cite{Q} and related  references.
Also,
the fact that every unital qubit channnel can be written as 
the average of $k$ unitary channels, where $k$ is the rank of the Choi matrix,
can be restated as: The mixed-unitary rank of unital qubit channel is always
the same as the Choi rank; see \cite{Girard}.
It would be interesting to identify other quantum channels with this property.

There is an one-one correspondence between 
affine maps on $\IR^3$ and trace preserving linear maps 
on $M_2$ that preserve Hermitian matrices. 
Specifically, an affine map 
 $\phi: \IR^3 \rightarrow \IR^3$  defined by
$$\phi(a,b,c) = (a,b,c)A + (a_0,b_0,c_0)$$
for a real matrix $A \in M_3$ and a fixed vector $(a_0,b_0,c_0) \in \IR^3$
corresponds to the linear map $\Phi: M_2 \rightarrow M_2$ satisfying 
$ \Phi(I_2) = I_2 + a_0 X+ b_0 Y + c_0Z$ and 
$$
\Phi(aX+bY+cZ) = \hat a X + \hat b Y + \hat c Z 
\quad \hbox{ with } \quad (\hat a, \hat b, \hat c) = 
(a,b,c)A.$$
Clearly, the map $\Phi$ is unital if and only if $(a_0,b_0,c_0) = (0,0,0)$.
In \cite{RS}, the authors investigated the effect of the affine map 
$\phi: \IR^3 \rightarrow  \IR^3$ induced by a general qubit channel 
$\Phi: M_2 \rightarrow M_2$ on the Bloch sphere. 
It would be interesting to identify simple conditions on a real matrix $A \in M_3$ 
and $(a_0, b_0, c_0) \in \IR^3$ that guarantee the affine map $(a,b,c) \mapsto (a,b,c)A + 
(a_0, b_0, c_0)$ corresponds to a qubit channel.  Some analysis of this problem
has been done in \cite{RS}.

\medskip\noindent
{\large \bf Acknowledgements}

\medskip
Li is an affiliate member of the Institute for Quantum Computing, University of Waterloo.
His  research  was  supported by the Simons Foundation Grant  851334. 
The authors gratefully acknowledge the valuable comments provided by the referees. 
They also like to express their gratitude to Professor Alexander M\"{u}ller-Hermes, 
Professor Frederik vom Ende, and Professor Seok-Hyung Lie for their 
helpful suggestions.


\begin{thebibliography}{WW}

\bibitem{BK}
J. Bae1 and L.-C. Kwek (2015),
Quantum state discrimination and its applications,
Journal of Physics A: Mathematical and Theoretical  48, 083001.


\bibitem{Caban}
P. Caban, J. Rembieli\'{n}ski, K.A. Smoli\'{n}ski, Z. Walczak (2015), 
Classification of two-qubit states,
Quantum Inf Process 14, pp.\ 4665–4690.

\bibitem{Choi} M.D. Choi (1975), Completely positive linear maps on complex matrices,
Linear Algebra Appl. 10, pp.\ 285-290.

\bibitem{CW}
M.D. Choi and P.Y. Wu (1990), Convex combinations of projections, Linear Algebra 
Appl. 136, pp.\ 25-42.

\bibitem{Girard} M. Girard, D. Leung, J. Levick, C.K. Li, V. Paulsen, Y.T. Poon,
and  J. Watrous(2022),
On the mixed-unitary rank of quantum channels,
Comm. Math. Phys. 394 no. 2, pp.\ 919-951.

\bibitem{GIN}
L. Gyongyosi, S. Imre and H. V. Nguyen (2018), A Survey on Quantum Channel Capacities,
IEEE Communications Surveys \& Tutorials 20, pp.\ 1149-1205.
 
 \bibitem{LS}
 L. Landau and R. Streater (1993), On Birkhoff's theorem for doubly stochastic 
 completely positive maps of matrix algebras, 
 Linear algebra Appl. 193, pp.\ 107–127.
 

\bibitem{MP}
A. M\"{u}ller-Hermes and C. Perry (2019), All unital qubit channels are 4-noisy operations,
Letters in Mathematical Physics 109, pp.\ 1-9.

\bibitem{NC}
M.A. Nielsen and I.L. Chuang (2004), 
Quantum Computation and Quantum Information. Cambridge University Press.

\bibitem{Aron}
A. Pasieka, D.W. Kribs, R. Laflamme, and R. Pereira (2009), 
On the geometric interpretation of single qubit quantum operations 
on the Bloch sphere,
Acta Applicandae Mathematicae 108, Article number: 697.

\bibitem{Q}
N. Quesada, A. Al-Qasimi and D.F.V. James (2012),
Quantum properties and dynamics of $X$ states,
Journal of Modern Optics 59, pp.\ 1322-1329.

\bibitem{RS} M. B. Ruskai, S. Szarek (2002),
E. Werner, An analysis of completely positive
trace-preserving maps on $M_2$, 
Linear Algebra Appl. 347, pp.\ 159-187.

\bibitem{S} I. 
Schur (1923),  \"{U}ber eine Klasse von 
Mittelbildungen mit Anwendungen auf die Determinantentheorie, Sitzungsber. 
Berl. Math. Ges. 22, pp.\ 9–20.


\bibitem{VV} F. Verstraete and H. Verschelde (2022), 
On quantum channels, arXiv:quant-ph/0202124

\bibitem{W}
M.M. Wilde (2017), 
Quantum Information Theory (2nd ed.), Cambridge University Press.
\end{thebibliography}
\end{document}